\documentclass[letterpaper]{article}
\usepackage{aaai}
\usepackage{times}
\usepackage{helvet}
\usepackage{courier}

\usepackage{amsmath, amsthm, amssymb}
\usepackage{tabularx}
\usepackage{graphicx}
\usepackage{multirow}
\usepackage{subfigure}
\usepackage{url}
\usepackage{rotating}

\begin{document}

\title{People are Strange when you're a Stranger:\\Impact and Influence of Bots on Social Networks}
\author{Luca Maria Aiello, Martina Deplano, Rossano Schifanella, Giancarlo Ruffo \\ Computer Science Department\\ Universit\`a degli Studi di Torino \\ Torino, Italy}
\maketitle

\begin{abstract}

Bots are, for many Web and social media users, the source of many dangerous attacks or the carrier of unwanted messages, such as spam. Nevertheless, crawlers and software agents are a precious tool for analysts, and they are continuously executed to collect data or to test distributed applications. 
However, no one knows which is the real potential of a bot whose purpose is to control a community, to manipulate consensus, or to influence user behavior. It is commonly believed that the better an agent simulates human behavior in a social network, the more it can succeed to generate an impact in that community. 
We contribute to shed light on this issue through an online social experiment aimed to study to what extent a bot with no trust, no profile, and no aims to reproduce human behavior, can become popular and influential in a social media. Results show that a basic social probing activity can be used to acquire social relevance on the network and that the so-acquired popularity can be effectively leveraged to drive users in their social connectivity choices. We also register that our bot activity unveiled hidden social polarization patterns in the community and triggered an emotional response of individuals that brings to light subtle privacy hazards perceived by the user base.

\end{abstract}

\section{Introduction}\label{sec:introduction}

%

The capillary diffusion of online social media and the growing number of mobile devices connected to the Internet have determined an unprecedented expansion of the pervasiveness of the Web in real life. Online and offline spheres are today strictly entangled and individual and collective human behavior is increasingly predictable from, or even influenced by, the dynamics of the digital world.

The pivotal role that social media had in coordinating the protesters in the 2011 Arab Spring wave~\cite{panisson11slashdot}, the effectiveness of online communication on political opinion formation~\cite{conover11predicting}, or the rapidity of spreading of earthquakes alarms through social media~\cite{sakaki10earthquake} are clear examples of the direct effect that the Web can have on human lives.

Mining the user-generated data streams from social media allows the analysts to detect, better understand and even try to predict real world events such as the spreading of diseases~\cite{culotta10towards}, the mood of the inhabitants of urban areas~\cite{quercia11mood}, the trends in the stock market~\cite{bollen11twitter}, or even the political election results~\cite{tumasjan10predicting,metaxas11predict}. The social and economic relevance of such opportunities has increased the interest not only in learning online social dynamics more in depth but also in understanding the extent to which such dynamics can be manipulated and controlled. For instance, Twitter has been leveraged to unfairly boost the consensus on some politicians for electoral purposes by automatically generating high amounts of fake tweets supporting particular opinions~\cite{ratkiewicz11detecting}. 


Efforts have been spent in developing bot-detection techniques~\cite{lee10uncovering}, but the real potential of robots in manipulating online social networks is still widely unknown. We contribute to the exploration of this field by investigating the relation between \textit{trust}, \textit{popularity} and \textit{influence}, three building blocks on which dynamics of opinion formation and information spreading on social substrates are based. Differently from previous data-driven studies on this matter, we go beyond the exploratory data analysis and we set up a social experiment based on a \textit{bot} interacting with human users in a online social environment. Our goal is to understand to what extent a bot with no public information can become popular and influential or, more in general, how it can impact the dynamics of the network.

Surprisingly, our experiments reveal that an untrustworthy individual can become very relevant and influential through very simple automated activity. Moreover, the social reactions aroused by the bot's activity give interesting insights on the possibility of automated attacks to affect the structure of the social substrate. 

\subsection{Contributions and roadmap}

Obtained results 1) provide explanations on the roots of \textit{popularity} and \textit{influence} in a social media context, 2) give useful insights about how a social \textit{recommender} (or \textit{spammer}) can be extremely effective, 3) provide a direct test of accuracy of modern \textit{recommendation} techniques on a real user base, and 4) show how the intervention of an external, anomalous entity in a social network can alter its dynamics and unveil \textit{social polarization} phenomena. Moreover, differently from traditional data-driven analysis that must face the hard task of inferring causal relationship between facts just by observing their temporal sequence, we can directly explore the causality of events by acquiring the perspective of an agent that interacts with other users. We make available upon request the anonymized data collected during and right after the experiment.


In Section~\ref{sec:dataset} we describe the main features of the social media on which we based our experiment and we present the structure of the bot we used to interact with other users. In Sections~\ref{sec:popularity} and~\ref{sec:influence} we explain the results obtained by the bot in terms of gain of popularity and influence, and in Section~\ref{sec:sentiment} we overview the side effects of its activity in terms of altering the dynamics of interaction on the social network.


\section{Dataset and experimental setup}\label{sec:dataset}

\begin{table}[tp]
	\small
	\centering
	\begin{tabular}{ccc|cc}
		& \multicolumn{2}{c|}{Jul 2010} & \multicolumn{2}{c}{Nov 2011} \\
		& \textbf{Social} & \textbf{Comm} & \textbf{Social} & \textbf{Comm}\\
		\hline
		Nodes & 86,800 & 72,054 & 179,653 & 95,121 \\ 
		
		Links & 697,910 & 483,151 & 1,566,369 & 677,652 \\
		
		$\left\langle k_{out}\right\rangle$ & 8.0 & 6.71 & 8.72 & 7.12 \\ 
		
		Reciprocation & 0.57 & 0.59 & 0.51 & 0.60 \\
		
		GSCC size & 62,195 & 33,700 & 127,997 & 44,072 \\ 
		\hline
	\end{tabular}
	\caption{Basic structural quantities in the aNobii social and communication networks at the beginning and at the end of the experiment. $\left\langle k_{out}\right\rangle$ is the average out-degree, reciprocation is the ratio of bidirectional arcs, and GSCC is the greatest strongly connected component.}
	\label{tab:anobii_evolution_stats}
\end{table}

Our experiment has been set up on \url{aNobii.com}, a social network for book lovers. Besides specifying classic profile information such as age and geographic location, aNobii users can fill a personal digital \textit{library} with the titles they have read and they can enrich them with ratings, annotations and reviews. Users socialize by means of thematic groups and by establishing pairwise social connections. Social ties are directed and can be created without any consent of the linked user. Two types of mutually-exclusive social links are available, namely \textit{friendship} and \textit{neighborhood}, but their behavior is identical, therefore in the following we consider a single \textit{social} network composed by their union. Every profile page contains a public \textit{shoutbox}, where any user can write messages. Message exchange implies the existence of another social graph that we call \textit{communication} network, where a directed arc $A \rightarrow B$ models a message from $A$ to $B$'s shoutbox. The arcs are weighted with the cumulative number of messages. Even though a few privacy settings are available, the whole profile information is usually public.

\begin{figure}[tp]
\centering
\includegraphics[width=.99\columnwidth]{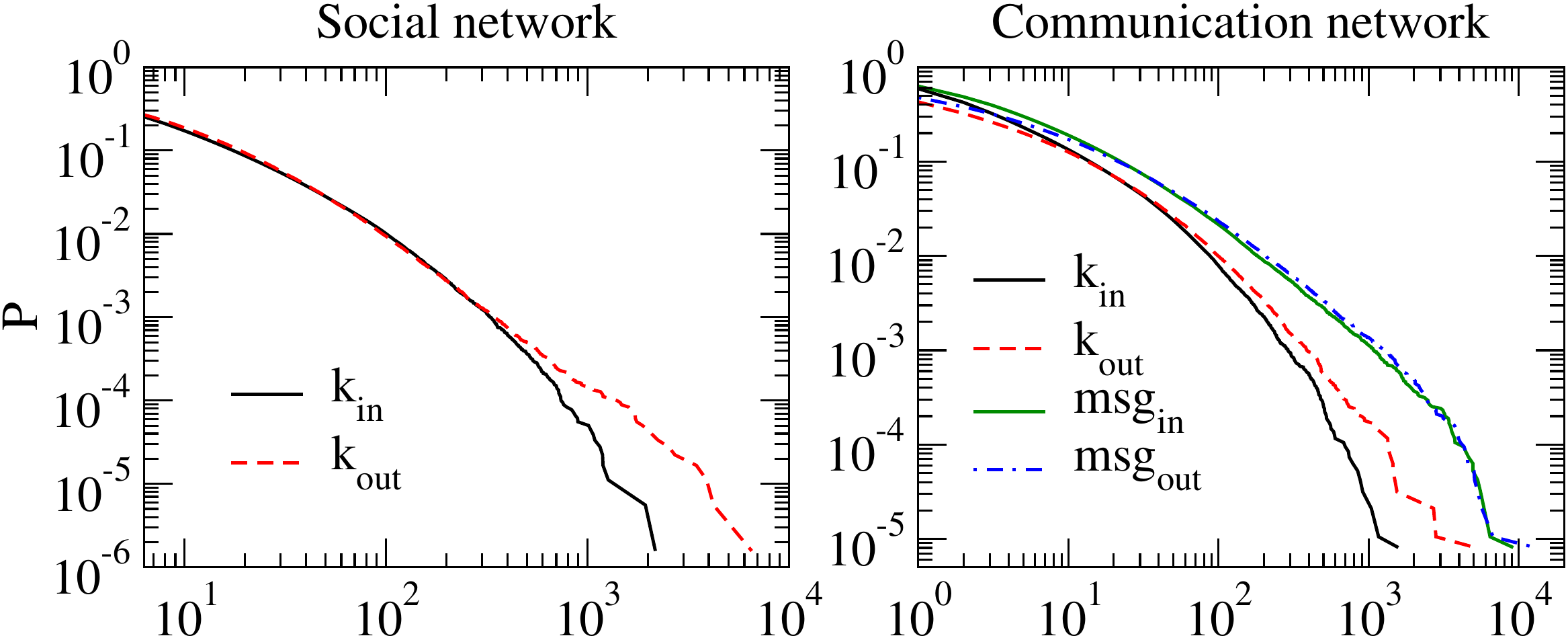}
\caption{Complementary cumulative density functions of in- and out-degree $k_{in/out}$ in the social and communication graphs and of in- and out-weighted degree $msg_{in/out}$ (i.e., overall number of received/sent messages) in the communication graph at the end of year 2011.}
\label{fig:distributions}
\end{figure}

Table~\ref{tab:anobii_evolution_stats} reports basic structural statistics of social and communication graphs at the beginning and at the end of the 1.5 years time frame of our experiment. The network roughly doubled its size in that period and kept a high reciprocation trend in social linking and a rather low average degree if compared to other general-purpose social media~\cite{ugander11anatomy}. However, not surprisingly, the social activity in terms of social linking and message exchange is rather broad, as shown by Figure~\ref{fig:distributions}.

We started collecting data in September 2009 through a periodic activity of Web crawling to study the dynamics of link creation in the network. The relatively small size of the graph allowed us to crawl all the nodes reachable by a BFS strategy once every 15 days (the crawled data are available upon request). We were constrained to create an empty stub account to access some protected sections of user profiles accessible only to subscribed users, and use its credentials to access such pages. The profile was automatically named \textit{lajello} after the email prefix of the subscriber.

Our experiment started with an accidental discovery. In July 2010 the default user settings of the website silently changed so that every visit of a logged user automatically started leaving a trace in a private \textit{guestbook} of the visited profile. The users can check their personal guestbook and see the last $30$ visitors of their libraries. As a result, our crawler left a trace of its passage in all the profiles reached by the BFS approximatively twice a month. 

\begin{figure}[tp]
\centering
\includegraphics[width=.99\columnwidth]{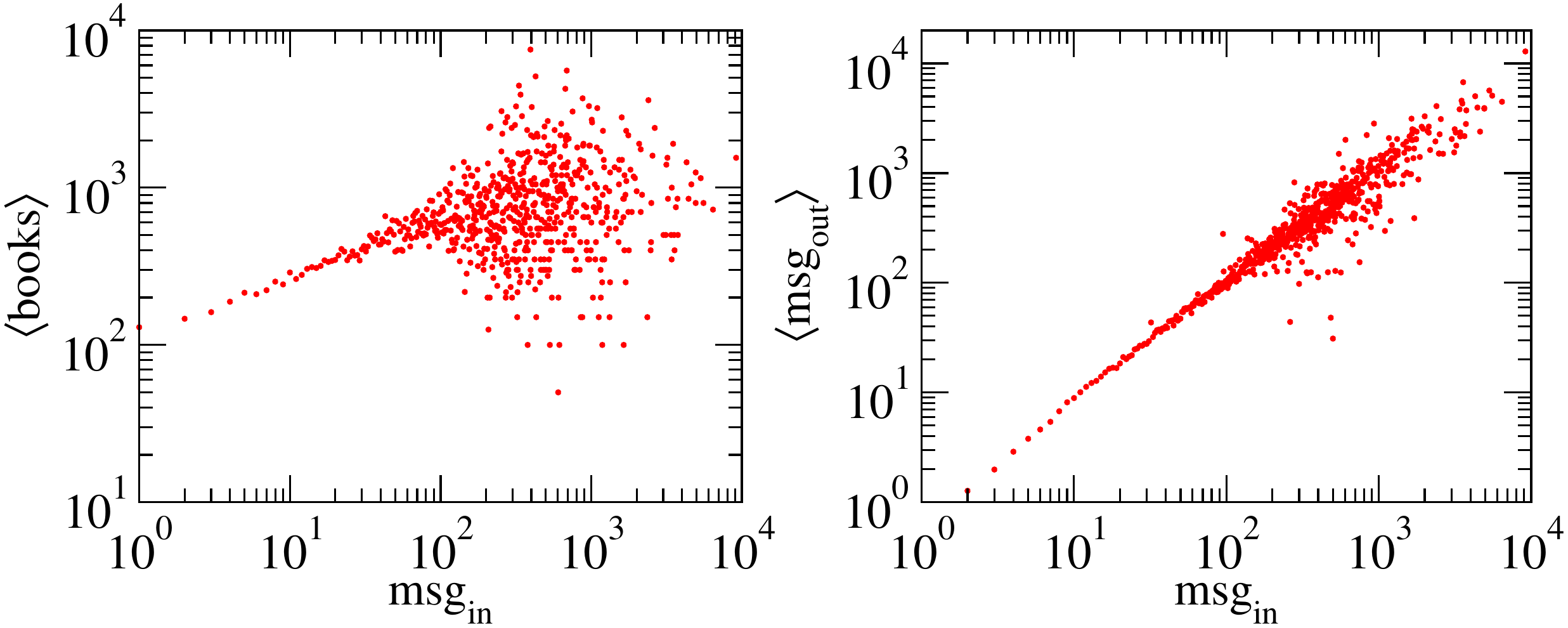}
\caption{Average number of books and sent messages for users who received the same number of messages $msg_{in}$.}
\label{fig:correlations}
\end{figure}

\begin{figure*}[tp]
\centering
\includegraphics[width=1.8\columnwidth]{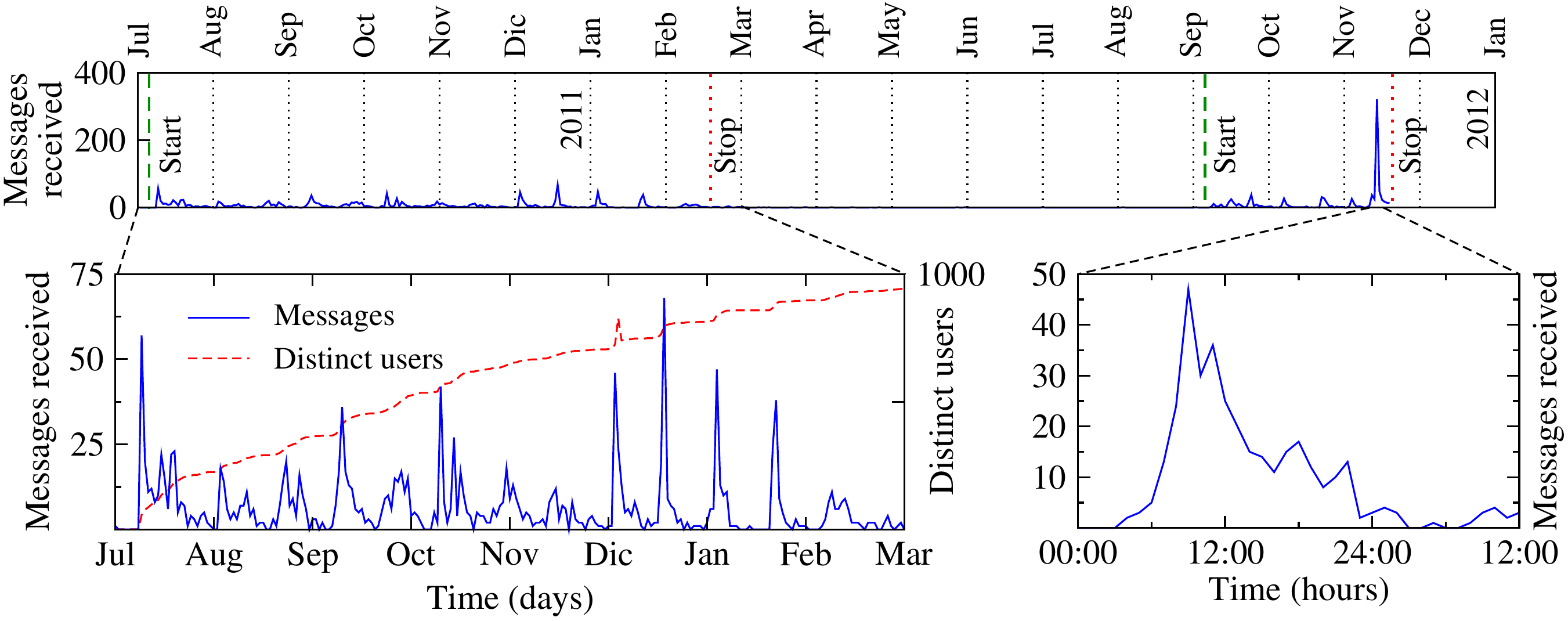}
\caption{User response to \textit{lajello}'s probing activity taking place between Jul 2010 and Jan 2012. The timeline at the top represents the overall histogram of the number of received messages on a daily basis during the whole period. The times of start and stop of the periodic crawling activity are marked. Bottom-left plot shows a finer-grain histogram for the period from Jul 2010 to Mar 2011. The cumulative number of distinct users writing messages is reported as well. Bottom-right plot shows a zoomed view on the 36 hours after the first recommendation message sent by the bot (see Section~\ref{sec:influence}).}
\label{fig:lajello_popularity}
\end{figure*}

The unexpected reactions the bot caused by its visits motivated us to set up a social experiments in two parts to answer the question \textit{``can an individual with no trust gain popularity and influence?''}. In the first part we kept running the periodic crawling to check the variation over time of the user response to the bot visits. In the latter phase we run a modified version of our bot that was able to communicate with other users, asking them to take some action. Results are discussed in Sections 3 and 4.

\section{Gain of popularity}\label{sec:popularity}

The popularity of an individual in social media is widely considered to be strictly related to the amount of incoming connections or to the number of received messages, meaning that a popular user is characterized by a network structural position at the center of star-shaped clusters~\cite{stoica10star}. Researchers have long investigated the defining traits of the popular users in social media. Factors determining user popularity range from the amount and the kind of profile features exposed to the public~\cite{lampe07familiar,strufe10profile} to the extroverted attitude of individuals in social communication~\cite{quercia12personality}. More in general, the creation of social trust bonds is strongly correlated with the similarity of opinions and profiles~\cite{golbeck09trust}.

From a quantitative point of view, also in aNobii we find a clear positive correlation between the number of incoming messages with the amount of books publicly shown in the library and with the amount of sent messages, as shown in Figure~\ref{fig:correlations}. This is not surprising given the tendency to reciprocity of social protocols and considering that aNobii is focused on books only, but still it gives an hint about the importance of being \textit{active} to gain the attention of other participants. However, identifying the main cause of popularity among many user activity indicators that are all strictly correlated one to another is not trivial. In this perspective, our social experiment is aimed to isolate a single, minimal social activity (namely the visit to other profile pages) and to test its effectiveness in gaining popularity.

\begin{figure}[tp]
\centering
\includegraphics[width=.99\columnwidth]{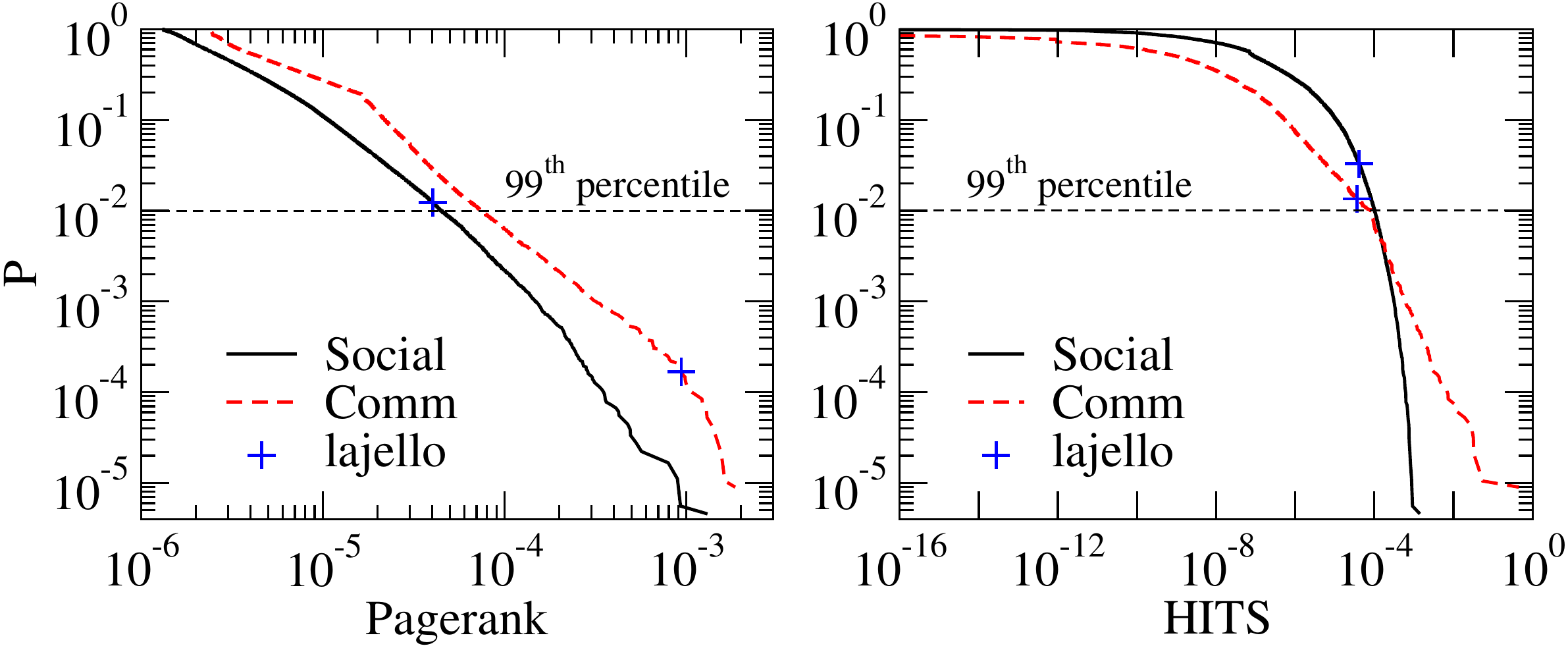}
\caption{Complementary cumulative density functions of Pagerank and HITS in social and communication networks. A guide to the eye marks the upper bound of the $99^{th}$ percentile. The scores of \textit{lajello} are marked with blue crosses and belong, from left to right in the figure, to the $98^{th}$, $99^{th}$, $98^{th}$, and $96^{th}$ percentiles.}
\label{fig:pagerank}
\end{figure}
%

As soon as the new aNobii settings on visits took place in July 2010, our bot started collecting a massive response from visited users, and every round of visits triggered a burst of comments on the bot's public wall. Figure~\ref{fig:lajello_popularity} shows the histogram of the number of visits in time. Higher spikes occurs right after the start of the two-monthly visits. The different height of spikes depends from the duration of the crawl, that has been inconsistent (from 24 hours to some days) due to the discontinuous uptime of the Web service. We also observe that the cumulative number of distinct users shouting grows constantly and roughly linearly in time.

Interestingly, the rapid bursts of messages happening after the visits do not trigger any self-feeding effect in the incoming communication stream. When the crawler terminates the visit round, people quickly stop sending messages except to respond with the same feedback when next round is performed. To confirm this, in mid February, after 15 crawling rounds, we stopped the bot until September 2011. A negligible amount of messages has been received in the inactivity period and the user feedback recovered the same pace when we resumed the crawling.

At the end of its activity, \textit{lajello}'s profile has become one of the most popular in the website, boasting 2,435 public messages from 1,263 different users, more than 200 private messages, more than 66,000 visits to its profile, and 211 incoming social connections (partitioned in 125 neighbors and 86 friends). Considering the very limited size of the social network and the range of the degree distributions of aNobii users (see Figure~\ref{fig:distributions}), such values are extremely high. In particular, in December 2011, the bot was the $2^{nd}$ user for the number of distinct people writing on its shoutbox (the first one is a well-known writer) and the $28^{th}$ in the number of received messages, with just 10 months of activity. Moreover, it was among the $0.3\%$ of users with the highest in-degree.

Even though the amount and weight of incoming connections is usually considered a good proxy for popularity, more sophisticated metrics taking into account the social status of the connected nodes could give a more accurate picture on the amount of popularity gained, following the intuition that an individual is more popular if it is followed by other popular people. To this end, we computed the distributions of Pagerank and HITS on social and communication networks (Figure~\ref{fig:pagerank}). Even considering such metrics, \textit{lajello} positions itself in the very tail of the distributions.

In short, our experiment gives strong support to the thesis that popularity can be gained just with continuous ``social probing''. This is particularly striking from the perspective of the security of online social environments since we have shown that a very simple \textit{spambot} can attract great interest even without emulating any aspects of a typical human behavior.

\section{Influence}\label{sec:influence}

Previous studies on the interdependence between popularity and influence in social media show that popularity does not imply the ability of being influent and vice-versa~\cite{cha10measuring}. It appears that influence depends on a combination of the position on the social graph~\cite{ilyas11} and of the language used in public communications~\cite{quercia11mood} and, intuitively, trust seems to be a precondition for both popularity and influence.

However, the great response collected from the bot with a very simple automated activity naturally arises the question about the possibility for an untrusted, empty profile to be influent on other community members.

\begin{table}[tp]
\centering
\small
\begin{tabularx}{235pt}{c|X}
\textbf{Feature} & \textbf{Description}\\
\hline
Common neighbors & $CN(u,v) = |\Gamma_{out}(u) \cap \Gamma_{in}(v)|$ \\
Triangle overlap & $\frac{CN(u,v)}{\Gamma_{out}(u)}$ \\
Resource allocation & $\frac{1}{k_{out}(u)} \sum_{z \in (\Gamma_{out}(u) \cap \Gamma_{in}(v))}\left(\frac{1}{k_{out}(z)}\right)$  \\
Reciprocation & Binary attribute, whether the inverse link $(v,u)$ is already present\\ 
Contact list & $\sigma_{sim}$ between social contacts vectors \\
Library & $\sigma_{sim}$ between library vectors  \\
Groups & $\sigma_{sim}$ between group vectors  \\
Group size & Size of the smallest group the two users have in common \\
\hline
\end{tabularx}
\caption{Features used in the recommendation of a directed link between users $u$ and $v$, along with their description. $\Gamma_{in/out}(u)$ is the set of $u$'s in/out neighbors, $k_{out}(u) = |\Gamma_{out}(u)|$. Resource allocation has been defined by Zhou et al.~\citeyear{zhou09predicting}. $\sigma_{sim}$ denotes the cosine similarity.}
\label{tab:features}
\end{table}

To answer this question we arrange a second phase of the social experiment in which the bot tries to convince users to perform an action. Many possible activities are allowed by the aNobii service, but we focus on the task of persuading users to add a new neighbor in their contact list. This choice is motivated by two reasons mainly. First, recommending a social activity rather than one that you can carry on individually (such as the adoption of a new book) has a higher potential of triggering some interaction between the users that are subjected to the experiment. This is desired since we are also interested in studying the social and emotional reactions of the user base to the bot's activity. Second, we observe that despite the vast literature on link recommendation in social networks~\cite{lu11link}, as far as we know there are no reported results on the actual effectiveness of the most recent suggestion techniques on actual users. Our experiment is structured such that the validity of modern contact recommendation techniques and the bot's degree of influence can simultaneously validated.

\begin{figure}[tp]
\centering
\includegraphics[width=0.90\columnwidth]{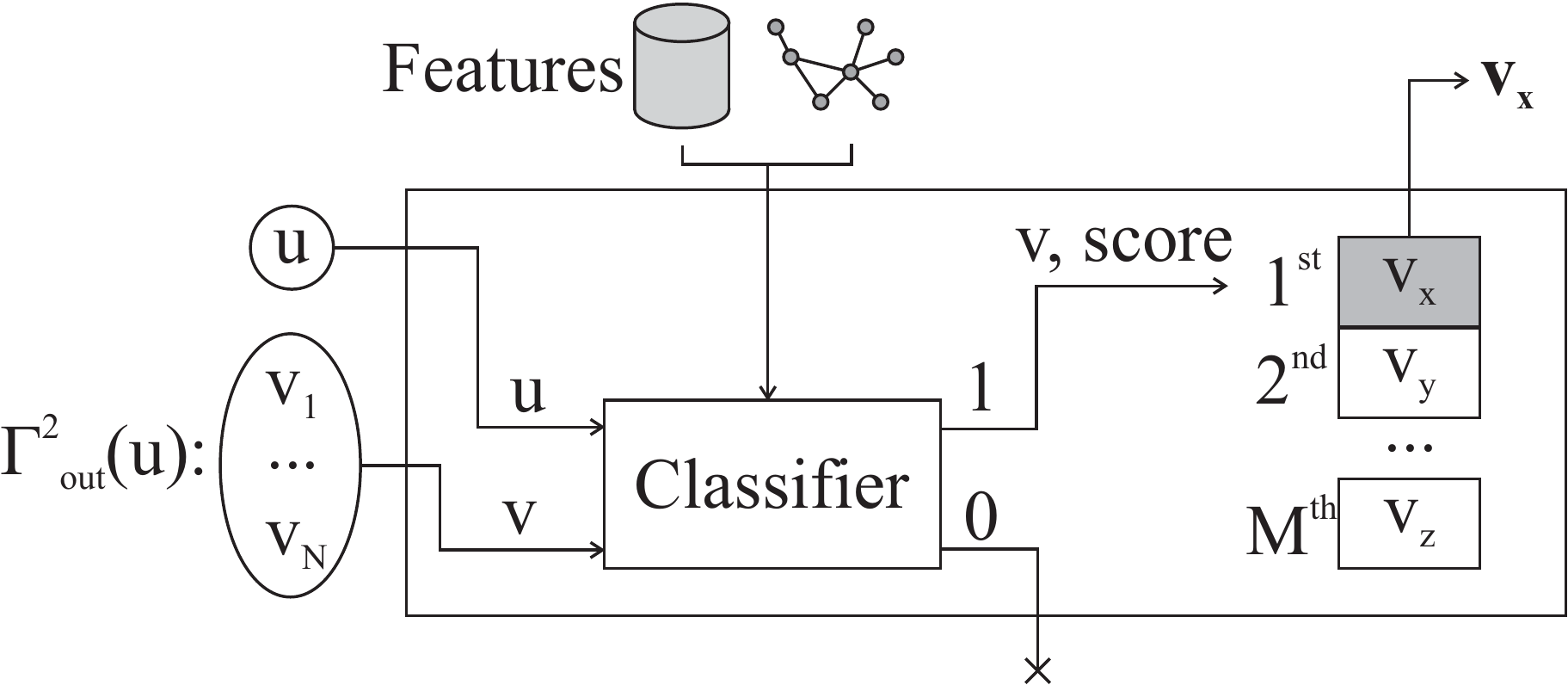}
\caption{Logical scheme of our recommender system. The target user $u$ together with its distance-2 neighbors and the respective features are given to the classifier. The best match is then given in output.}
\label{fig:recommender_scheme}
\end{figure}

First, let us introduce our link recommendation method. Based on the findings of previous work on link prediction and recommendation~\cite{aiello11friendship} we adopt a machine-learning approach that combines features extracted from the profile pages and structural features defining the position of users on the social graph. Predictive metrics usually exploit the observation that the process of link creation in social networks is driven by proximity and homophily~\cite{aiello10socialcom} i.e., people tend to close social triangles with people more similar to them.

We use the Rotation Forest classifier ---which proved to be the best performing for this task among all WEKA classifiers~\cite{weka09}--- trained on the feature set listed in Table~\ref{tab:features}. Among a larger set of features, we found those to be the most predictive through a Chi Squared feature selection. To first test the effectiveness of the classifier we used a balanced training set of $20k$ pairs of users residing 2-hops away in the social graph. Positive examples are pairs that got connected with a social tie between September and November 2010 and negatives are chosen randomly among those who did not. On a 10-fold cross validation we obtained $0.82$ as Area Under the ROC Curve (AUC), which denotes a good quality of the selected predictors.

The trained classifier can be easily used to produce recommendations on the fly. Given a user $u$, the classifier is applied between $u$ and all her distance-2 neighbors $\Gamma^{2}_{out}$. All the contacts in $\Gamma^{2}_{out}$ for which the classifier gives a positive response are sorted by the classifier's confidence score and the contact at the top of the list is selected as the one to recommend. The logical scheme of the recommender is depicted in Figure~\ref{fig:recommender_scheme}. We consider only distance-2 neighbors to keep the computational complexity low, but this strategy does not worsen much the quality of the recommendation given that in the most frequent case links are created between distance-2 pairs~\cite{aiello10socialcom}.

Once our recommender has been set, we used it to perform direct recommendations to the aNobii users. Differently from the first phase of the experiment, we focused on users who declared Italian nationality on their profile. This is motivated by the fact that $60\%$ of the aNobii users are from Italy and they represent the most active core of the network.

We structured the experiment as follows. First, we listed all the Italian users that wrote at least one message on the bot's shoutbox (we call them \textit{followers}) and an equal number of those who did not (\textit{non-followers}) taken at random among all the other Italian users with at least 10 books in their library, to avoid inactive users. Then we randomly sampled $50\%$ of users from this pool and produced for them a single contact suggestion using our recommender. For the remaining $50\%$ we produced random recommendations. Finally, we randomly sampled $25\%$ of the recommendation pairs $(u,v)$ (where $u$ is the user to be recommended and $v$ the recommended contact) and we produced the corresponding reciprocal recommendations $(v,u)$. At the end of this procedure we obtained five equal-size recommendation categories (followers recommendation, followers random, non-followers recommendation, non-followers random, and reciprocal) for a total of approximatively 3,000 different recommendations of non-existent social ties. 

Using a random message generator that has been written specifically for this experiment, we produced a message in Italian language for each user in the recommendation sample. Messages were all different for number and type of sentences used and for the use of lexical classes but all of them contained a suggestion for the recipient to add a certain user to her neighbors list.

\begin{figure}[tp]
\centering
\includegraphics[width=0.9\columnwidth]{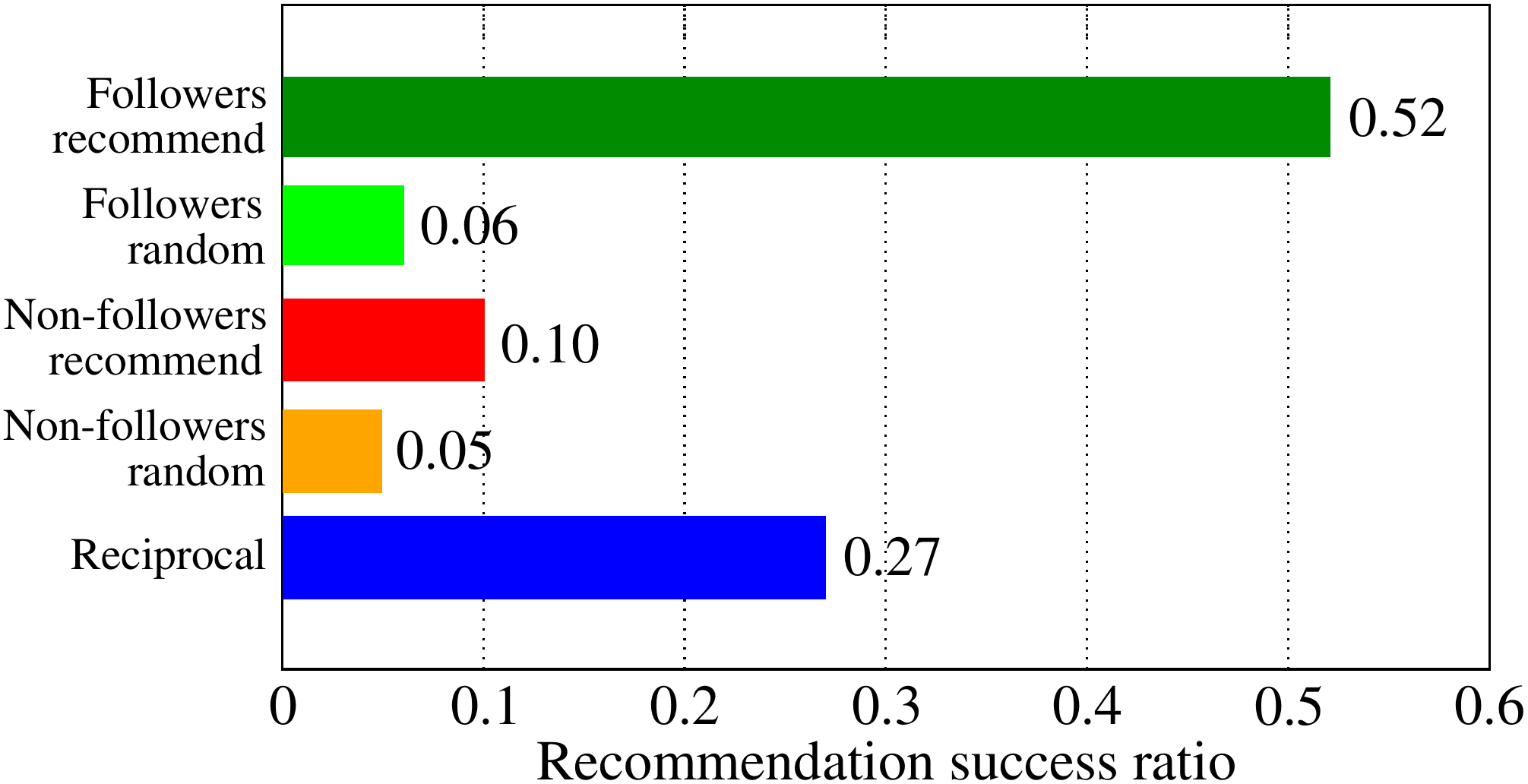}
\caption{Distribution of successful contact suggestions among the different recommendation classes. ``Followers'' denote the users who wrote at least one message to \textit{lajello}. ``Recommend'' and ``random'' indicate whether the recommendation has been performed using our classifier or just picking a random user. ``Reciprocal'' class includes recommendations of links that reciprocates other recommendations taken at random from any of the previous classes.}
\label{fig:lajello_recommendation_hist}
\end{figure}

We sent all the recommendation messages to the public shoutboxes of target users during the night of November $8^{th}$ 2011. From the early morning after we registered a bursty user response. In the 36 hours following the experiment, more than 400 shouts where published on \textit{lajello}'s shoutbox. Histogram in Figure~\ref{fig:lajello_popularity} shows how the peak of messages received during the day after is roughly one order of magnitude higher than any other peak observed during the past crawling activity.

An even more surprising feedback was given by the link recommendation results. Among the 361 users who created at least one social connection in the 36 hours after the recommendation, the $52\%$ followed the suggestion given by the bot. The distribution of the successful suggestions over the different categories (Figure~\ref{fig:lajello_recommendation_hist}) reveals three interesting patterns.

First, non-random recommendations dispensed to followers have been far more effective than those given to non-followers, in other words, \textit{lajello} has a greater persuasive power over those who are more aware of its presence and activity. Second, the effectiveness of thoughtful suggestions based on our recommender is at least double than random recommendations, meaning that users are mainly reluctant to obey blindly (even though a non-negligible $11\%$ of the successful recommendation belongs to the random categories). Finally, it is surprising to see that recommendations are more likely to succeed when both directions of the link are suggested to the two endpoints. This is mainly due to the fact that bidirectional recommendations are more likely to trigger some communication between the two endpoints that in many cases ends up in the agreement to form a new social tie.

Shortly after registering these results, the bot's account was suspended by the system administrators and all the recommendation messages were canceled. Even though we could not register the feedback of all the recommended users, the obtained results still lend themselves to meaningful interpretations from two different points of view.

From a social network analysis perspective, we observe that untrustworthy individuals in social media can become influent if they have enough popularity, which can be gained with automated activity. Even if our experiment has been forced to be limited in time and restricted to a relatively small number of users, its outcomes give a very strong evidence about the extent to which a community could be manipulated by external agents. We consider this a surprisingly outbreaking result in this field.

From the service provider angle instead, we learn that a \textit{social recommender} can be very effective ($52\%$ of success ratio in our case). Even if fair comparisons across different recommendation techniques and datasets are difficult, our results appear to be even better than recent results on link recommendation based on temporal datasets~\cite{backstrom11supervised}. This suggests that many recommendations that are labeled as false positives in such data-driven studies may result to be successful when they are actually proposed to the end user. Unfortunately, the same argument can be applied to malicious attackers: external unauthorized robots programmed to pilot user opinions and actions can be extremely effective in a social media context. For this reason, additional care should be put in the design of security devices of social media to avoid such hazards.

\section{Emotional reactions and social polarization}\label{sec:sentiment}

\begin{figure}[tp]
\centering
\includegraphics[width=0.9\columnwidth]{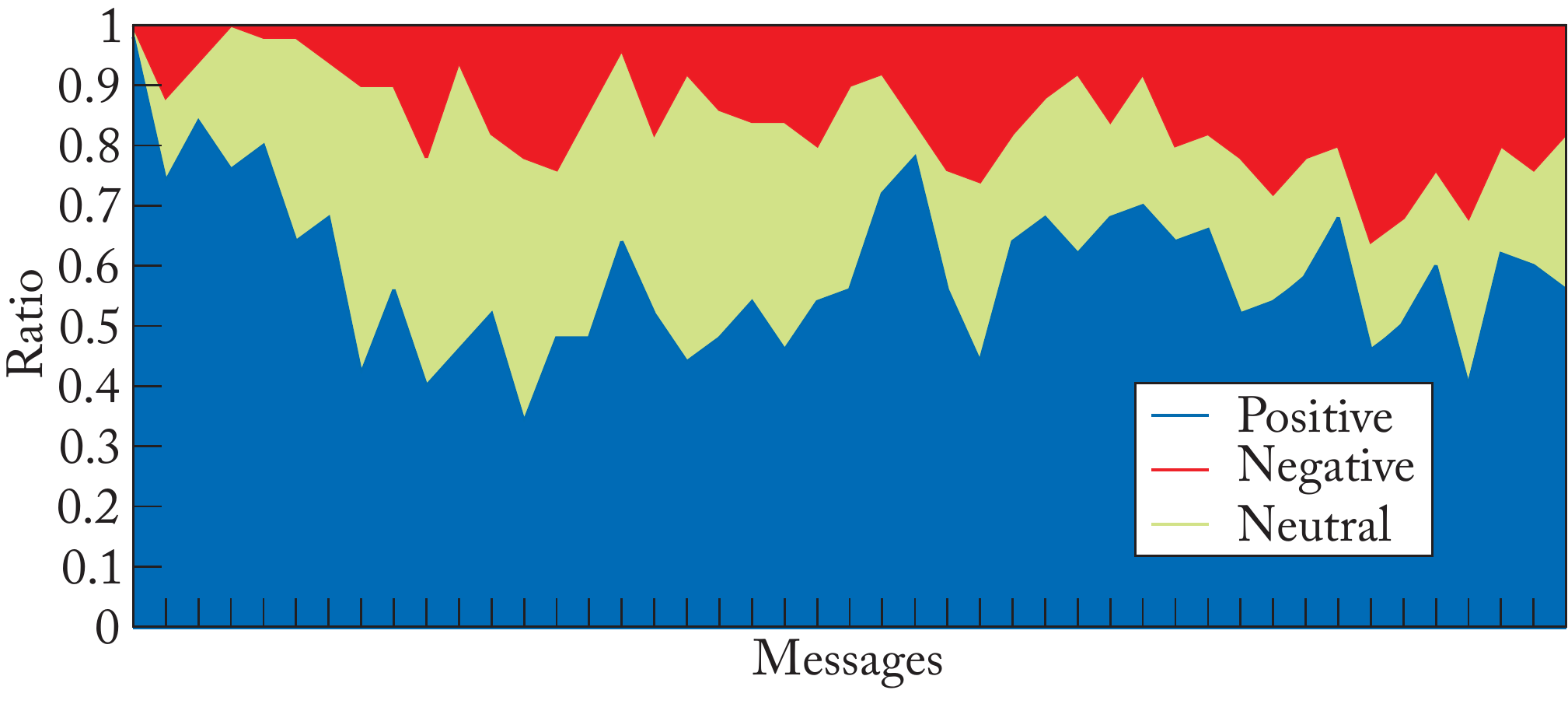}
\caption{Stacked area graph depicting the ratio of messages with positive (bottom blue), negative (top red), or neutral (center green) connotation in \textit{lajello}'s shoutbox. On every point on the x-axis we present the statistic over 50 consecutive messages.}
\label{fig:sentiment}
\end{figure}

In previous Sections we focused on the quantitative analysis of the user response to our bot to assess its popularity and influence across the community, but an additional relevant outcome of the experiment pertains the qualitative sphere of the collected user reactions.

The most striking result that is perceivable at a first glance is the wide variety of sentiments expressed in the messages received by \textit{lajello}, ranging from veneration to hatred. This has been allowed by the fact that many people believed the bot to be a human user and only in the latest phase of the experiment rumors about the possibility of \textit{lajello} being a robot began to slowly spread over the network. This result is coherent with another recent ``socialbot'' experiment on Facebook~\cite{boshmaf11socialbot}, in which $80\%$ of unaware users accepted the friendship request of a bot, as well as with previous experiments on social phishing~\cite{jagatic07social}.

Moreover, many users revealed deep discomfort for the fact that an unknown user was periodically visiting their profiles, even if our bot explored just the information available to anyone else in the community. This empirically confirms previous studies on the perception of privacy of users on online social media~\cite{levin09notions} showing that users are very concerned about the privacy risks given by the requests of \textit{strangers} without a good reputation or with no profile picture. Moreover, another explanation for such emotional reactions is given by the \textit{fear of control}, widely discussed by Michel Foucault~(\citeyear{foucault75discipline}), who exemplifies this phenomenon using the paradigm of the \textit{Panopticon}, the architectural model of a structure in which people actions are constantly observed and recorded. Several users openly expressed in their messages the fear of control or the perception of a strong privacy violation.



A fine-grain text analysis that is able to capture all the sentiment nuances expressed by the users goes beyond the scope of this work. However, to assess the amount of users that were positively or negatively impressed by the activity of the bot, we manually annotated all the received messages with a ternary label \textit{positive/neutral/negative}. All the annotations have been performed by three authors and a majority vote has been applied to pick the final label. Just a few ambiguous cases occurred. On average, we found that $59\%$ of messages have a positive connotation and $16\%$ are negative instead. By expanding this aggregate data on a temporal scale we note interesting fluctuations in the trend of consensus and dissensus, as shown in Figure~\ref{fig:sentiment}. 

\begin{figure}[tp]
\centering
\subfigure[Social network]{
\label{fig:polarization1}
\includegraphics[width=0.50\columnwidth]{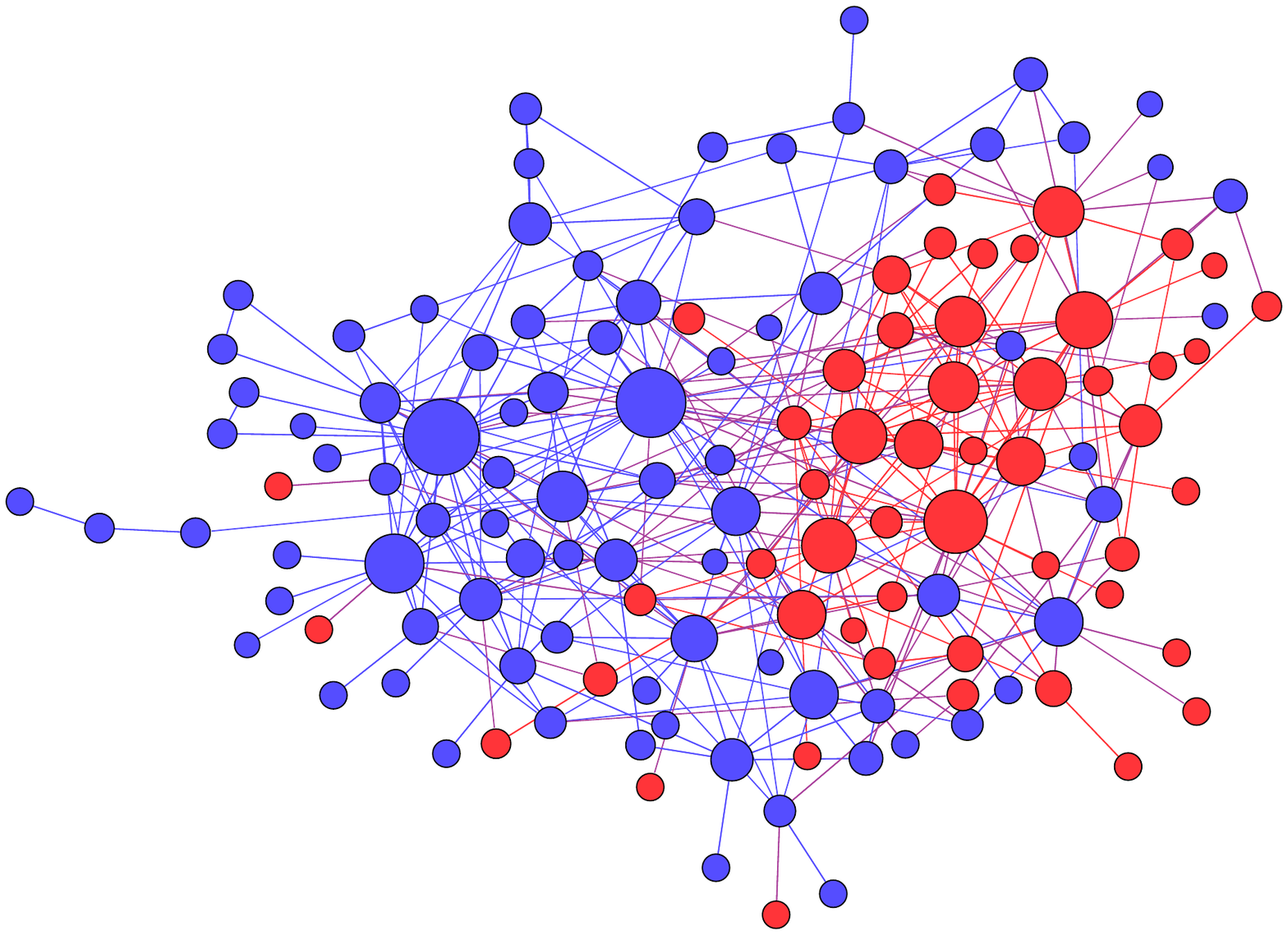}
}
\subfigure[Communication network]{
\label{fig:polarization2}
\includegraphics[width=0.40\columnwidth]{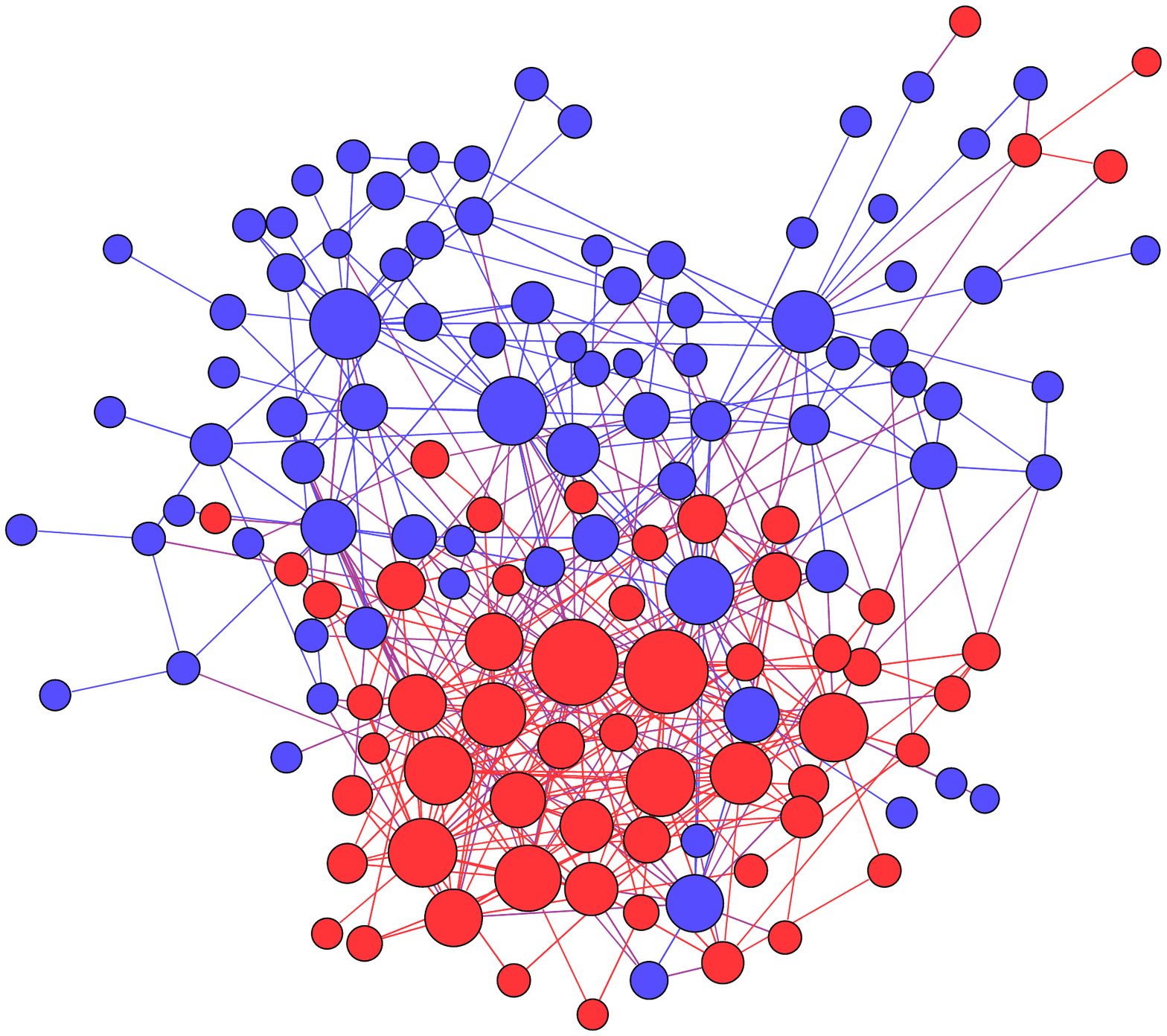}
}      
\caption{Networks induced by the members of the two groups opposing on \textit{lajello} topic. Blue and red circles represent users supporting and criticizing the bot, respectively. Node size is proportional to the degree. Singleton nodes, directionality and weight of edges are not shown for simplicity.}
\label{fig:polarization}
\end{figure}

\begin{table}[tp]
	\small
	\centering
	\begin{tabular}{ccc|cc}
		& \multicolumn{2}{c|}{\textbf{Actual}} & \multicolumn{2}{c}{\textbf{Randomized}} \\
		& Intra & Inter & Intra & Inter\\
		\hline
		Social 			& $74\%$ & $26\%$ & $52\%$ & $48\%$ \\
		Communication	& $71\%$ & $29\%$ & $49\%$ & $51\%$ \\
		\hline
	\end{tabular}
	\caption{Portion of links that reside inside a cluster (intra) or, conversely, that connect users belonging to different clusters (inter) in the real networks and in their randomized versions.}
	\label{tab:polarization}
\end{table}

Initially, users notice for the first time the visits of the bot and reply giving it a warm welcome. Soon after, the unknown reason of the visits and the permanently empty profile start raising some uncertainty in the community, and people begin asking for explanations but without expressing an explicit sentiment. Meanwhile, a small but growing percentage of hostile messages were addressed to the bot. The number of neutral messages gets thinner in time, even if the positive mood remains always prevalent. The most evident polarization of opinions occurs in the last $400$ messages, after the recommendation, when people started discussing whether the behavior of \textit{lajello} could have been acceptable. This phenomenon is coherent with studies on social polarization on Twitter~\cite{yardi10dynamic}, where user interactions on controversial topics reinforce the separation of opposing opinions.

Polarization in user opinions could be also tracked from the analysis of thematic groups. We counted seven groups that have been spontaneously dedicated to \textit{lajello} since its appearance. The main threads of discussion were focused on exchanging opinions about its identity, but also strong judgments about its activity have been expressed. In particular, right after receiving the recommendations from the bot, two opposing groups where quickly created: one in favor of \textit{lajello}, named ``Nessuno tocchi lajello'' (tr.``Hands off lajello'') currently with 102 members, and one against it with 72 participants, named ``I consigli (non richiesti) di lajello'' (tr. ``The (non-requested) suggestions of lajello'') . 

The sharp difference of opinions in the two groups motivated us to investigate how much the two factions are clustered when considering their social connections. Figure~\ref{fig:polarization} depicts the social and communication networks induced by the members of the two groups, displayed according to a force-directed layout~\cite{jacomy11forceatlas}. A first visual inspection already reveals that the two node categories tend to form two distinguishable clusters, even in a state of high link density.

To verify the presence of separate clusters more precisely, we perform a randomization test. We count the ratio between edges that fall within the same cluster (also called \textit{intra}-cluster edges) and those that connects nodes in different clusters (\textit{inter}-cluster edges) and we compare the obtained values with the same measurement performed on a randomized version of the network where nodes keep their out-degree but their links are rewired at random. Results in Table~\ref{tab:polarization} reveal a much stronger tendency of nodes to be connected inside their own cluster than in a random model.

An additional evidence of user polarization into two opinion clusters can be given by applying community detection algorithms to the networks and check whether the detected clusters match the group membership of nodes~\cite{conover11political}. We perform this test using the OSLOM community detection algorithm~\cite{lancichinetti11oslom} due to its good performance over other state of the art network clustering techniques. Since OSLOM outputs more than two clusters and we are interested in a binary polarization, we greedily merge together clusters whose majority of nodes belong to the same group, until we obtain two macro-clusters.

To assess the adherence of clusters to groups we compute the average fraction of correctly classified vertices~\cite{girvan02community} over 1000 realizations of OSLOM with random seed. We find that $80\%$ and $72\%$ of nodes in the social and communication networks respectively are correctly classified. This confirms the intuition that the emergence of opposed factions was rooted in the structure of social connections, but has been hidden until the system perturbation caused by \textit{lajello}. As shown in classical studies on social networks~\cite{zachary77information}, we can interpret this result also as the possibility of disrupting the network along the edges of detectable clusters, in a \textit{divide-et-impera} logic, by introducing a perturbating element in the community.

\begin{figure}[tp]
\centering
\includegraphics[width=0.95\columnwidth]{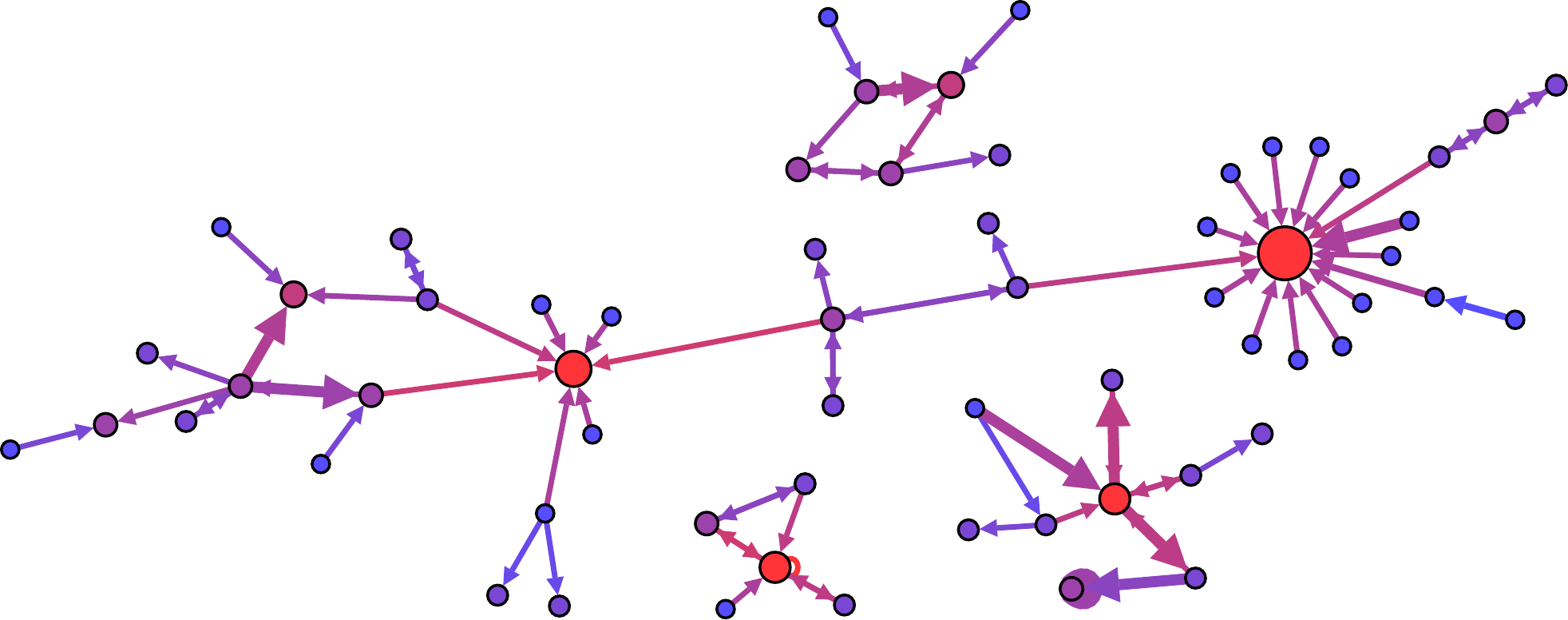}
\caption{Small subsample of communication network where edges represent messages containing the keyword ``\textit{lajello}''. Nodes are resized and colored according to their degree. The big red node on the right is a user who is emulating the behavior of \textit{lajello} to become popular. In this snapshot it is starting to attract the interest of other users who are writing on its shoutbox.}
\label{fig:lajello_meme}
\end{figure}

Finally, besides motivating the creation of new groups, the bot activity triggered also the creation of new communication links and new nodes in the network. Users started to talk about \textit{lajello} not only through the mentioned groups but also posting messages on each other shoutboxes. Tracking the keyword ``\textit{lajello}'' we can extract a subset of the communication network that maps the interaction channels over which our bot has been mentioned. The analysis of such filtered network reveals the presence of some nodes with an in-degree sensibly higher than the average, like the one shown in Figure~\ref{fig:lajello_meme}. Such nodes are \textit{clones} of our bot, i.e., empty profiles created to silently visit other users in the same way \textit{lajello} does. The activity of the clones soon attracted the attention of other users, who started writing messages on their walls mentioning \textit{lajello}.

\section{Conclusions}\label{sec:conclusions}

We presented the outcome of a social experiment aimed to explore the relation between \textit{trust}, \textit{popularity} and \textit{influence} in the dynamics of online social media. We showed that popularity in social networks does not require peculiar user features or actions, since an automated agent can acquire a surprisingly high popularity just by reiterating a simple activity of ``social probing''.

In a second phase of the experiment we sent friendship suggestions from our bot to a pool of users, providing random suggestions to some and thoughtful recommendations to others. As evidence that an untrustworthy user can be very influent if popular enough, we found that people more aware of the presence of the bot have been more inclined to follow its suggestions. Moreover, unlike previous work in the field, we validate the effectiveness of modern link recommendation techniques with an explicit feedback of the user base.

Last, we registered profound social dynamics alterations. Specifically, our bot was widely mistaken as a human user and its activity triggered the creation of new groups and communication channels, originated emulous users, raised explicit concerns about privacy violations and aroused a diffused fear of being controlled. Moreover, it unveiled preexistent latent social network clusters by triggering the polarization of opinions of community members.

Our results contribute to shed light on the effects that of automated agents can have in online social environments. Our findings have important implications on the way in which we conceive security and privacy in social networks. Since sociological and emotional sides of the experiment are crucial to investigate this issues, we plan for the future a more systematic sentiment analysis of the messages originated by the users in the wake of the bot's activity.

\section*{Acknowledgments}
This work has been partly supported by the INCA project. We are grateful to aNobii for making its data available, to Giuseppe Tipaldo, Alain Barrat, Ciro Cattuto and to the members of the NaN group at Indiana University for helpful discussions. Graphs visualization are made with Gephi~\cite{bastian09gephi} (\url{gephi.org}).

\bibliography{anobii_lajello_short}
\bibliographystyle{aaai}

\end{document}